# Interpretable Ransomware Detection Using Hybrid Large Language Models: A Comparative Analysis of BERT, RoBERTa, and DeBERTa Through LIME and SHAP

**Elodie Mutombo Ngoie,** University of Pretoria, Department of Computer Science, South Africa, u22608754@tuks.co.za

**Mike Nkongolo Wa Nkongolo**, University of Pretoria, Department of Informatics, South Africa, mike.wankongolo@up.ac.za

**Peace Azugo**, University of Pretoria, Department of Informatics, South Africa, peace.azugo@tuks.co.za

**Mahmut Tokmak**, Burdur Mehmet Akif Ersoy University, School of Applied Technology and Management, TÜRKİYE, mahmuttokmak@mehmetakif.edu.tr

## Abstract

Ransomware continues to evolve in complexity, making early and explainable detection a critical requirement for modern cybersecurity systems. This study presents a comparative analysis of three Transformer-based Large Language Models (LLMs) (*BERT, RoBERTa, and DeBERTa*) for ransomware detection using two structured datasets: UGRansome and Process Memory (PM). Since LLMs are primarily designed for natural language processing (NLP), numerical and categorical ransomware features were transformed into textual sequences using *KBinsDiscretizer* and token-based encoding. This enabled the models to learn behavioural patterns from system activity and network traffic through contextual embeddings. The models were fine-tuned on approximately 2,500 labelled samples and evaluated using accuracy, F1 score, and ROC-AUC. To ensure transparent decision-making in this high-stakes domain, two explainable AI techniques (*LIME and SHAP*) were applied to interpret feature contributions. The results show that the models learn distinct ransomware-related cues: BERT relies heavily on dominant file-operation features, RoBERTa demonstrates balanced reliance on network and financial signals, while DeBERTa exhibits strong sensitivity to financial and network-traffic indicators. Visualisation of embeddings further reveals structural differences in token representation, with RoBERTa producing more isotropic embeddings and DeBERTa capturing highly directional, disentangled patterns. In general, RoBERTa achieved the strongest F1-score, while BERT yielded the highest ROC-AUC performance. The integration of LLMs with XAI provides a transparent framework capable of identifying feature-level evidence behind ransomware predictions. Hence, this work introduces a novel hybrid LLM-XAI methodology for interpretable ransomware detection, providing practical insights for cybersecurity analytics and supporting the design of trustworthy defensive systems.

## 1. Introduction

Cybersecurity threats such as ransomware have become increasingly sophisticated, exploiting vulnerabilities across networks and systems to encrypt user data and demand ransom payments (Kharraz et al., 2015). Detecting these threats early is essential to protecting both individual users and enterprise infrastructures (Scaife et al., 2016). To support this effort, this study utilises two ransomware-related datasets: the Process Memory (PM) dataset (Mokoma and Singh, 2025) and the UGRansome dataset (Yan et al. 2024), which contain numerical and categorical features describing system activity, network traffic, and process behaviour.[1] The goal of this study is to apply Transformer-based Large Language Models (LLMs), specifically BERT, RoBERTa, and DeBERTa, to detect ransomware behaviour from these datasets. The models are trained on labelled ransomware features, employing tokenization to represent transformed numerical variables and their contexts as distinct tokens. The models are further analysed using explainable AI techniques such as LIME and SHAP to better understand the decision-making process and ensure transparency in model predictions. LIME and SHAP are used to identify key features that have the most influence over ransomware classification decisions, whether they are tokenised inputs or embedding dimensions. As such, this research answers the research question *"How XAI and LLM techniques can be integrated to improve the interpretability of ransomware classification?"*. The primary purpose of this research is to create a hybrid framework that can assist in classification of ransomware in network traffic and process memory datasets (Figure 1). First, the dataset will have to be pre-processed by replacing all null or NaN values, and be converted into tokens using *KBinsDiscretizer* to convert them into understandable text for the models (Figure 1). In the experiments, pre-trained transformer models such as BERT, RoBERTa, and DeBERTa are implemented to help in ransomware classification (Figure 1). We then apply XAI techniques such as LIME and SHAP to interpret model predictions. This process aims to visualize and analyse the relationships between features, attention mechanisms, and model performance to derive cybersecurity insights, emphasising the importance of explainable AI in ransomware detection.

---

[1] https://www.kaggle.com/code/thashannaick/ransomware-detection-using-llm-and-xai-techniques

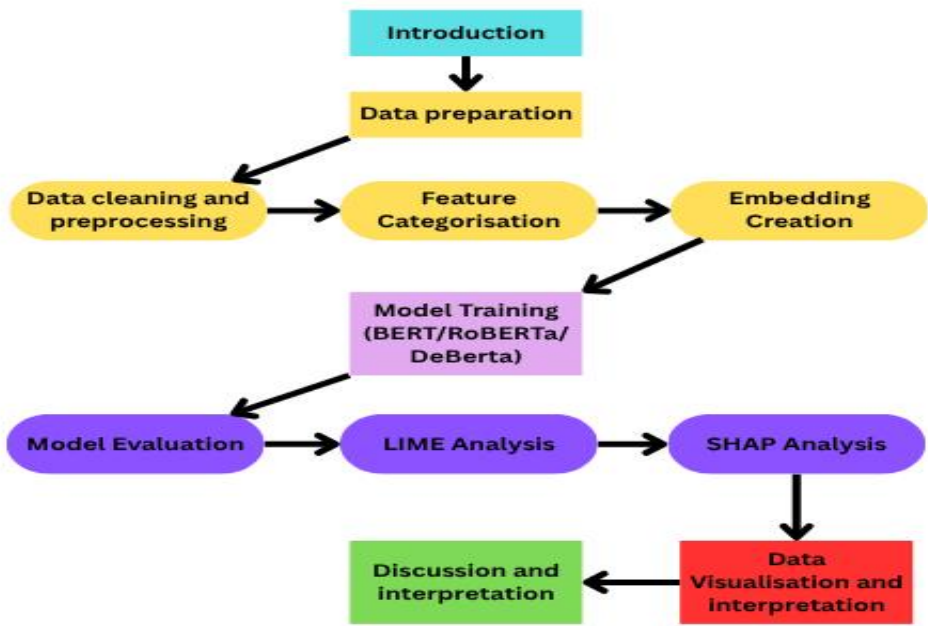

**Figure 1.** Computational framework

Unlike existing ransomware and intrusion-detection studies that primarily rely on classical machine-learning pipelines, handcrafted features, or conventional deep architectures (Ali et al., 2025) the methodology adopted in this study introduces a fundamentally different paradigm by leveraging Transformer-based Large Language Models (LLMs) and explainable AI. For example, Zhang et al. (2025) employ feature-selection heuristics, sensitivity analysis, and an optimized hybrid model to enhance ransomware detection, while Alkasassbeh et al. (2025) design a self-adaptive intrusion detection system using Deep Q-Networks to identify zero-day ransomware attacks. Patel et al. (2025) focus on securing wireless telesurgery networks through an intelligent rule-based and ML-driven framework, and Zeng et al. (2025) utilize a CNN-LSTM architecture to learn temporal malware patterns from sequential opcode data. Similarly, Su (2025) develops generative mathematical models enhanced by Chi-Square feature selection to improve predictive accuracy. In contrast, our study transforms structured ransomware telemetry into textual sequences using *KBinsDiscretizer* and token-based encoding, enabling BERT, RoBERTa, and DeBERTa to learn behavioural patterns through contextual embeddings, a methodological shift from tabular learning to language-model reasoning. Moreover, while prior studies rarely incorporate interpretability beyond sensitivity analysis, our work integrates LIME and SHAP to provide feature-level explanations, revealing how each LLM emphasizes distinct ransomware signals (file-operations, network behaviour, or financial indicators). This hybrid LLM-XAI methodology therefore advances the field by coupling modern language-model fine-tuning with transparent decision analytics (Jelodar et al., 2025) offering a more interpretable and behaviour-aware framework for ransomware detection than the predominantly feature-engineered or sequential-neural approaches found in related literature (Ali et al., 2025; Zhang et al., 2025; Patel et al., 2025).

**2. Data Preparation**

According to IBM (2023), data cleaning is the process of preparing raw data for analysis by removing or correcting inaccurate records, filling in missing values, and ensuring consistency across datasets. It is a critical phase in machine learning workflows, particularly for cybersecurity data, where noisy or incomplete records can lead to misleading model predictions and reduced threat-detection accuracy (Jain and Mitra, 2025). In this study, the cleaning process was applied to two ransomware-related datasets, namely: Process Memory (PM.csv) and UGRansome.csv, to prepare them for tokenization and training of Large Language Models (BERT, RoBERTa, and DeBERTa).

Recent studies have demonstrated that both the Process Memory (PM) and the UGRansome datasets have become widely adopted benchmarks for contemporary ransomware detection research. Mokoma and Singh (2025) utilized Process Memory features in *RanViz*, transforming API call sequences into time-series categorical representations to classify ransomware behaviours through visual and machine-learning techniques. Similarly, Rios-Ochoa et al. (2025) employed UGRansome in a comprehensive evaluation of dataset quality and model performance for ransomware family attribution, highlighting its suitability for benchmarking detection algorithms under realistic conditions. These works collectively demonstrate a growing trend in the cybersecurity community: Process Memory and UGRansome are now recognized as key modern datasets for modelling, analysing, and detecting ransomware activity, providing reliable foundations for machine-learning and behavioural-analysis experiments. A missing-value analysis was performed using "*pandas.isnull().sum()*" to identify incomplete fields in both datasets. There was a high percentage of missing values (about 50%) in six categorical columns, namely: ip_address, threats, flag, address, seed_address, protocol, and category. These null values were likely due to incomplete network logs. But since removing half of the dataset would harm the model generalisation, we handled the missing numeric values by replacing them with the median value of the dataset while replacing the categorical columns with the mode of the set (Figure 1). There were also label inconsistencies, the UGRansome dataset used a column called "prediction" for classification labels, while PM.csv used "label".

Preprocessing and encoding are essential steps in preparing data for machine learning and, in this case, for fine-tuning transformer-based Large Language Models (LLMs). These steps ensure that the raw numerical and categorical ransomware data are cleaned, standardized, and transformed into a format compatible with models such as BERT, RoBERTa, and DeBERTa, which are originally designed to process natural language rather than numerical input. Preprocessing improves data quality by handling inconsistencies such as missing or noisy values, ensuring that all features contribute meaningfully to model learning (Figure 2). Encoding, on the other hand, converts structured data into text-like tokens, enabling transformer models to understand and learn patterns from cybersecurity features as if they were linguistic sequences (Figure 2). Without these steps, LLMs would not be able to process the numerical attributes of ransomware datasets effectively, since they require sequential, and tokenized inputs (Jaffal et al., 2025). In this study, preprocessing began by filling missing numerical values with the median of each feature to maintain statistical consistency while avoiding bias from extreme values (Figure 2).

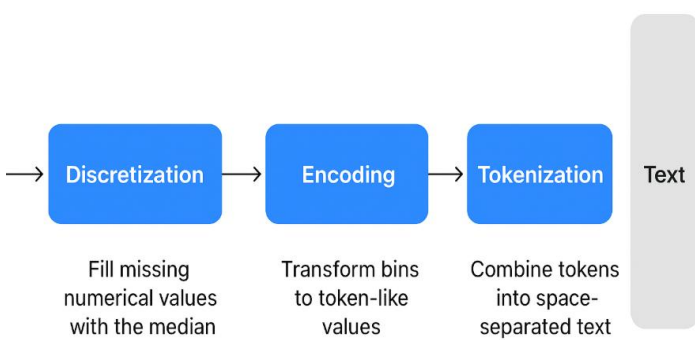

**Figure 2.** The proposed numerical-to-text tokenization pipeline (NTTP)

Then, discretization (binning) was applied using the *KBinsDiscretizer* function, which divided continuous numerical features into five quantile-based bins. Each bin was assigned an ordinal value representing its range, effectively converting continuous values into discrete categories. Following discretization, encoding transformed these binned values into token-like text representations (Figure 2). For example, a numeric column named cpu_usage with a bin value of 3 became the token "cpu_usage_bin_3". Each row of the dataset was then combined into a space-separated string, forming a "sentence" of tokens that describe the behaviour or state of a system instance. The result was a new column called text, representing each record as a textual sequence suitable for input into transformer models (see Appendix). Table 1 shows each step (and code), the category in which it falls and the explanation on why it was implemented.

| Step / Code Block | Category | Explanation |
| --- | --- | --- |
| *python data[numeric_cols] = data[numeric_cols].fillna(data[numeric_cols].median())* | **Preprocessing** | Handles missing numeric values by replacing them with the median to ensure numerical stability. |
| *encoder = KBinsDiscretizer(...); binned_data = encoder.fit_transform(data[numeric_cols])* | **Preprocessing (Transformation)** | Discretizes continuous numeric variables into 5 quantile-based bins to reduces skewness and normalizes feature distributions. |
| *tokenized_df = pd.DataFrame(binned_data, columns=numeric_cols)* | **Preprocessing → Intermediate Step** | Creates a Data Frame of the binned numerical data before encoding into text tokens. |
| *for col in tokenized_df.columns: tokenized_df[col] = col + "_bin_" + tokenized_df[col].astype(int).astype(str)* | **Encoding** | Converts numeric bins into word-like tokens (e.g., bytes_sent_bin_3, cpu_usage_bin_1). This transforms numeric data into textual format for LLM compatibility. |
| *data["text"] = tokenized_df.apply(lambda x: " ".join(x.values), axis=1)* | **Encoding** | Combines all feature tokens in each row into a space-separated sentence, effectively turning a record into an LLM-readable text sequence. |
| *final_df = data[["text", label_col, "source"]]* | **Encoding + Finalisation** | Prepares the final dataset structure (inputs and labels) for model training — the dataset is now in encoded textual form. |

**Table 1.** Data transformation steps

**3. Embeddings**

Embeddings are numerical representations of words/tokens which capture their meaning for the model to understand them mathematically (Kumar et al., 2025). They are important for models like BERT, RoBERTa, and DeBERTa as they cannot read words directly and need embeddings to find patterns (Figure 3). BERT's embeddings convey rich textual meaning, but are often anisotropic, meaning few dominant directions explain the variance (Ahanger et al., 2025). For example, tokens like "bin" appear often in clusters showing that the model has missed their common structure. The separability between tokens is limited, showing that while BERT encodes rich contextual features, however, it doesn't clearly separate classes, so a bit of post-processing can help make its representations more balanced (Figure 4). RoBERTa seems to have more evenly distributed embeddings due to its byte-pair encoding (BPE). This simply means tokens are more separated in space so it can tell contexts and sub-word nuances apart better than what BERT could do (Figure 5). This additional information helps in differentiating better between label related terms maintaining stronger token differentiation, thus increasing its overall performance in downstream classification and contextual representation tasks.

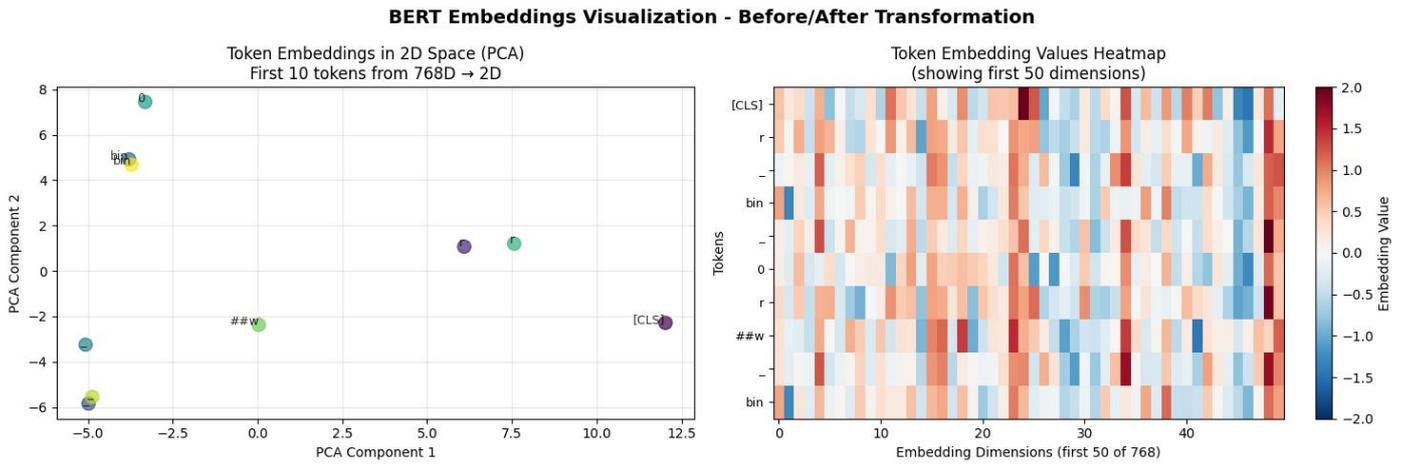

**Figure 3.** BERT token embeddings

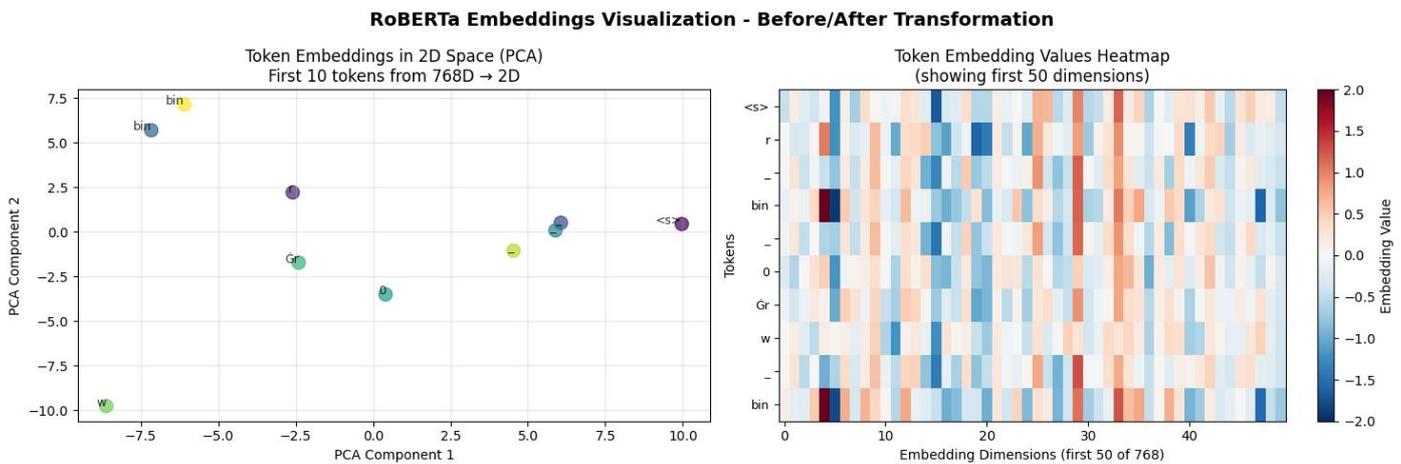

**Figure 4.** RoBERTa token embeddings

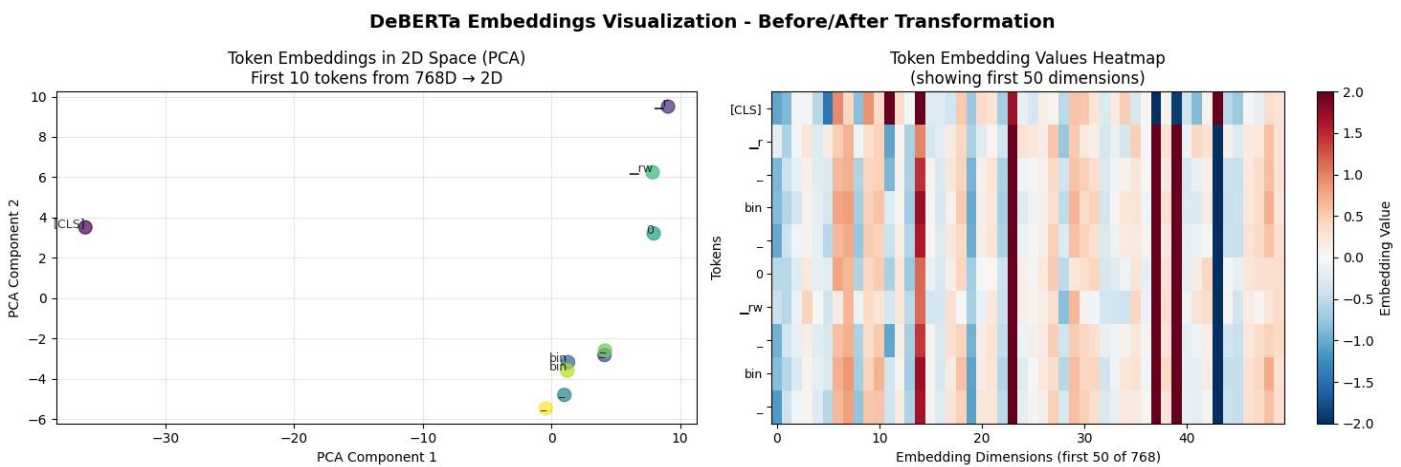

**Figure 5.** DeBERTa token embeddings

DeBERTa's embeddings seems more aligned due to its disentangled attention mechanism, separating word content from positional information (Cheng et al., 2025). This representation seems to yield more directions, which causes tokens to cluster along specific axes instead of spreading evenly like in BERT or RoBERTa. Since the model used is a smaller variant, it also has less parameters to distribute variance, which also contributes to the tighter, right aligned pattern. The dark vertical bands around dimensions 10, 20, and 40 show that embedding dimensions have higher activation where the model concentrates most of its variance, showing some shared structured features (Figure 3-5).

## 4. XAI Interpretability

LIME is a method that helps understand why a model made certain predictions (Lakshmi et al., 2025). It works by slightly changing input data and seeing how the model prediction changes. Average importance shows how much the model relies on certain tokens. Sign shows the direction of influence (Positive pushes the prediction towards the class like Ransomware), and small values mean weaker influence while larger ones mean the token had a strong effect on the models output (Figure 6).

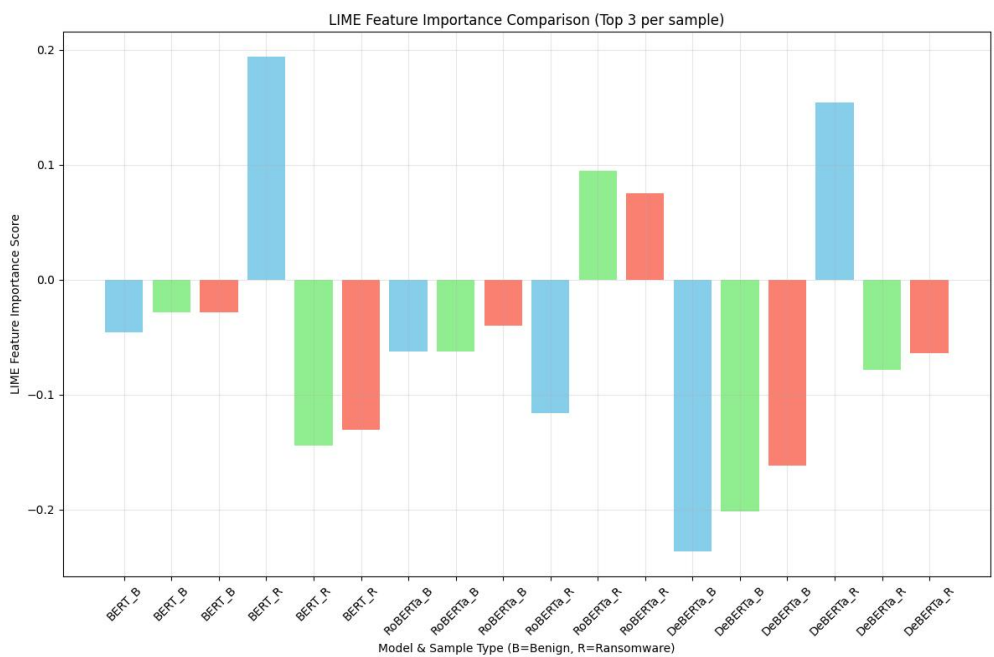

**Figure 6.** LIME interpretability

BERT shows the single strongest feature influence of any model (BERT_R). This large spike means BERT relies heavily on one specific clue to make its ransomware decision, this can lead to a high recall but potentially more false positives (Figure 6). RoBERTa's feature influences are more moderate and clustered, no single feature dominates. This balanced use of features explains the best F-1 score of 0.800 and shows the balance of precision and recall (Figure 6). DeBERTa shows the strongest negative influences, this means it places importance on all missing benign activities as evidence of ransomware (Figure 6). This approach contributes to a high precision of 0.837 as it requires specific conditions to be met first. The results are presented in Table 2.

| Model | Sample Type | Top Features & Scores | Interpretation & Rationale | Model Profile |
|---|---|---|---|---|
| **BERT** | **Ransomware** | port_bin_1= +0.1939 clusters_bin_0= -0.1444 rwxc_bin_0= -0.1307 | Heavily relies on a specific network port as a positive ransomware signal. Uses absence of normal cluster and file activity as supporting evidence. | Uses a clear strategy with one very strong positive feature. |
| | **Benign** | clusters_bin_0= -0.0458 netflow_bytes_bin_0= -0.0285 port_bin_0= -0.0285 | Identifies benign traffic by detecting low activity levels across network and port features. | |
| **RoBERTa** | **Ransomware** | clusters_bin_0= -0.1162 btc_bin_2= +0.0947 netflow_bytes_bin_2= +0.0455 | Combines absence of normal clusters with presence of Bitcoin activity (semantically meaningful) and specific network traffic patterns. | Uses a nuanced approach combining multiple feature types for robust decisions. |

| | Benign | clusters_bin_0= -0.0626<br>netflow_bytes_bin_0= -0.0625<br>port_bin_0= -0.0404 | Similar to BERT, identifies benign samples through low activity in standard network features. | |
| DeBERTa | Ransomware | rwxc_bin_0= -0.1769<br>clusters_bin_0= -0.1294<br>rw_bin_3= +0.0736 | Strongly weights absence of standard file operations while also detecting specific read-write patterns that match ransomware encryption behavior. | Focuses on core malicious behaviours (file operations) for high-confidence alerts. |
| | Benign | clusters_bin_0=-0.0779<br>netflow_bytes_bin_0= -0.0635<br>port_bin_0= -0.0546 | Consistent with other models, uses low network activity as indicator of benign traffic. | |
| *BERT* | | AVG importance | Benign – 0.0145 | Ransomware – 0.0888 |
| *RoBERTa* | | AVG importance | Benign – 0.0249 | Ransomware – 0.0583 |
| *DeBERTa* | | AVG importance | Benign – 0.0954 | Ransomware – 0.0532 |

**Table 2.** LLMs results

SHAP was used to help understand which tokens most influenced the models predictions, it worked by removing one token at a time and seeing how its prediction changes (Alvi et al., 2025). If removing a token mean the result significantly changed, the token was important. In our case positive values meant the token support the benign class, and negative values meant it supported ransomware. SHAP drew up horizontal bar charts to show the direction and magnitude of each feature's impact on the model's prediction. The key principle is

- *Bars extending to the RIGHT (Positive values) push the model toward predicting Ransomware.*
- *Bars extending to the LEFT (Negative values) push the model toward predicting Benign.*

For BERT there were 2 benign features that stood out by far, namely, cluster_bin_0 and netlfow_bytes_bin_0. This means it identifies normal activity by the absences of significant network activity. For BERT, the ransom features that heavily impacted decision making was anything with file operations (rw) combined with financial signals (usd). This also shows decisions are driven by few but very strong signals (Figure 7).

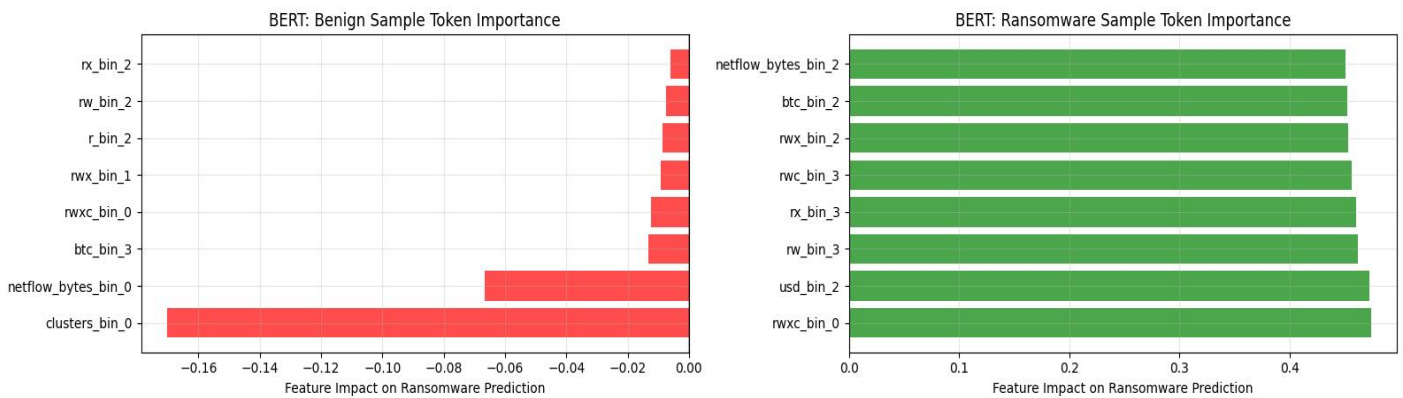

**Figure 7.** BERT feature importance

RoBERTa on the other hand used a simpler more focused rule by focusing on low network traffic to identify benign classification (Nandinee et al., 2025). For ransom classification, it used a wider spread of moderate-strength signals such as financial and file operation activities as the main indicators (Figure 8). Lastly, DeBERTa shows the strongest confidence in what constitutes benign activity, with the largest magnitude negative impacts. DeBERTa shows an extreme focus on financial transactions and is by far the strongest feature, while also relying on network traffic patterns (Figure 9). We observe that while the models share a common understanding of 'benign' activity such as low network traffic, they specialise in detecting different malicious patterns. BERT focuses on file operations, RoBERTa employs a balanced analysis including multiple financial markers, and DeBERTa specialises in financial and network signals (Table 3). This specialisation directly correlates with their performance metrics and provides actionable intelligence.

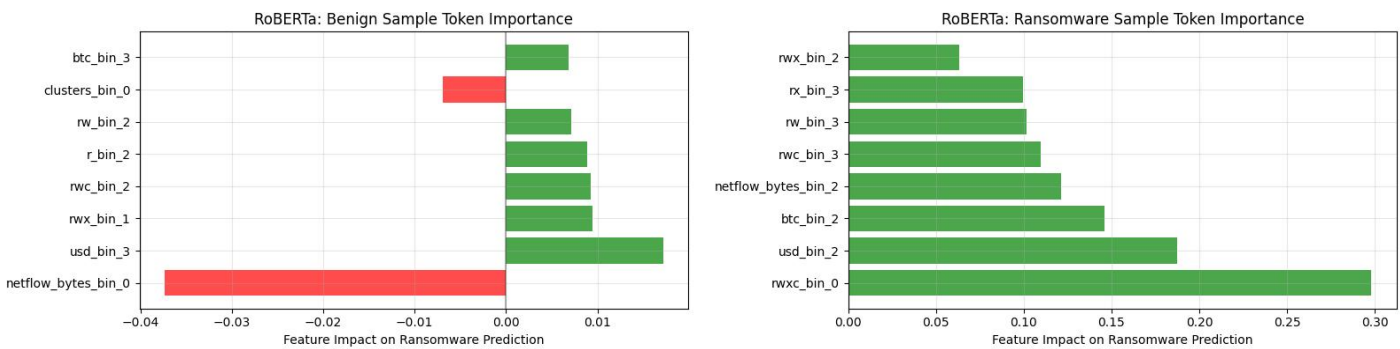

**Figure 8.** RoBERTa feature importance

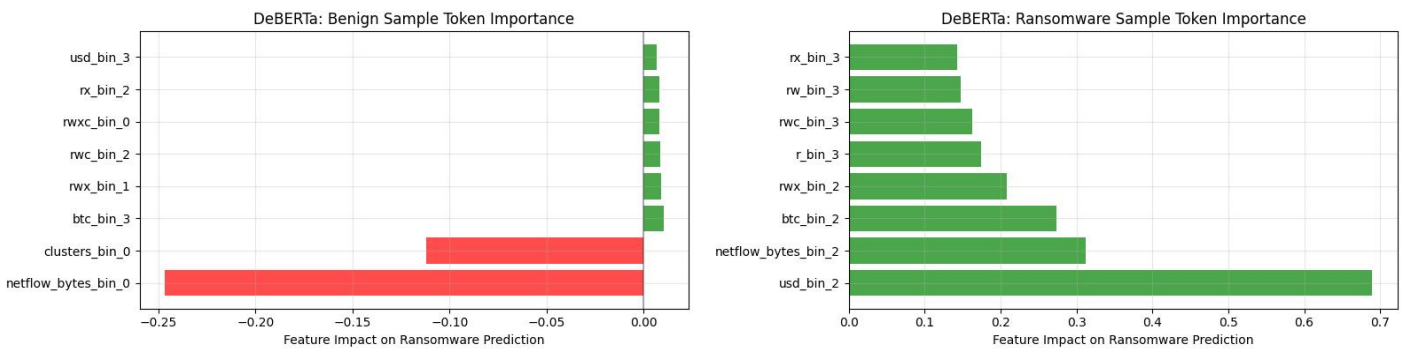

**Figure 9.** DeBERTa feature importance

| Model | Benign Avg Importance | Most Important Benign Feature | Ransomware Avg Importance | Most Important Ransomware Feature |
|---|---|---|---|---|
| **BERT** | 0.0304 | clusters_bin_0 | 0.4214 | rwxc_bin_0 |
| **RoBERTa** | 0.0116 | netflow_bytes_bin_0 | 0.1197 | rwxc_bin_0 |
| **DeBERTa** | 0.0420 | netflow_bytes_bin_0 | 0.2148 | usd_bin_2 |

**Table 3.** Model-wise feature importance analysis for benign and ransomware classes

**5. Data Visualisation**

This section presents the visual analysis of the UGRansome and Process Memory (PM) datasets, emphasizing feature distribution, skewness reduction, normalization, and comparative scaling. The goal is to identify key ransomware-related features and interpret their patterns across datasets. All numerical variables were examined for skewness to understand their data distribution before model training (Stratton et al., 2025). Several features showed significant right-skewed distributions, indicating the presence of extreme values and imbalance. Transformation techniques were applied to normalize these distributions, improving statistical symmetry and feature comparability. The UGRansome Transform Report shows that variables like btc (skew = 10.8160) and usd (skew = 3.2752) exhibited extreme skewness prior to transformation. Applying the Box-Cox method successfully reduced skewness to near-zero values (btc: -0.0234; usd: -0.1680). Similarly, network features such as clusters and netflow_bytes showed substantial improvement after transformation. The Process Memory (PM) Transform Report revealed similar outcomes. Memory-access features (rw, rwx, rwxc) were heavily skewed before normalization (rw: 15.4731; rwxc: 9.0549). After applying Box-Cox and Yeo-Johnson transformations, skewness was notably reduced (rw: 0.0798; rwx: -0.0483), resulting in more balanced distributions suitable for machine learning. We plotted 2 sets of Histograms (Figure 9), for all numerical features before and after transformation. The raw histograms displayed heavy right tails, while the transformed versions appeared more symmetric and centred around the mean. After applying the Min-Max scaling, the feature ranges were normalized between 0 and 1, enabling direct comparison across variables of different units and magnitudes. The histogram analysis highlights that normalization balances feature contributions, improving the fairness of feature weighting during ransomware classification. This process ensures consistent model interpretation across diverse operational metrics. Two correlation heatmaps was generated to visualise relationships among normalised features between the sets for UGRansome (Figure 10) and Process Memory (Figure 11).

**5.1. Histogram of feature distributions**

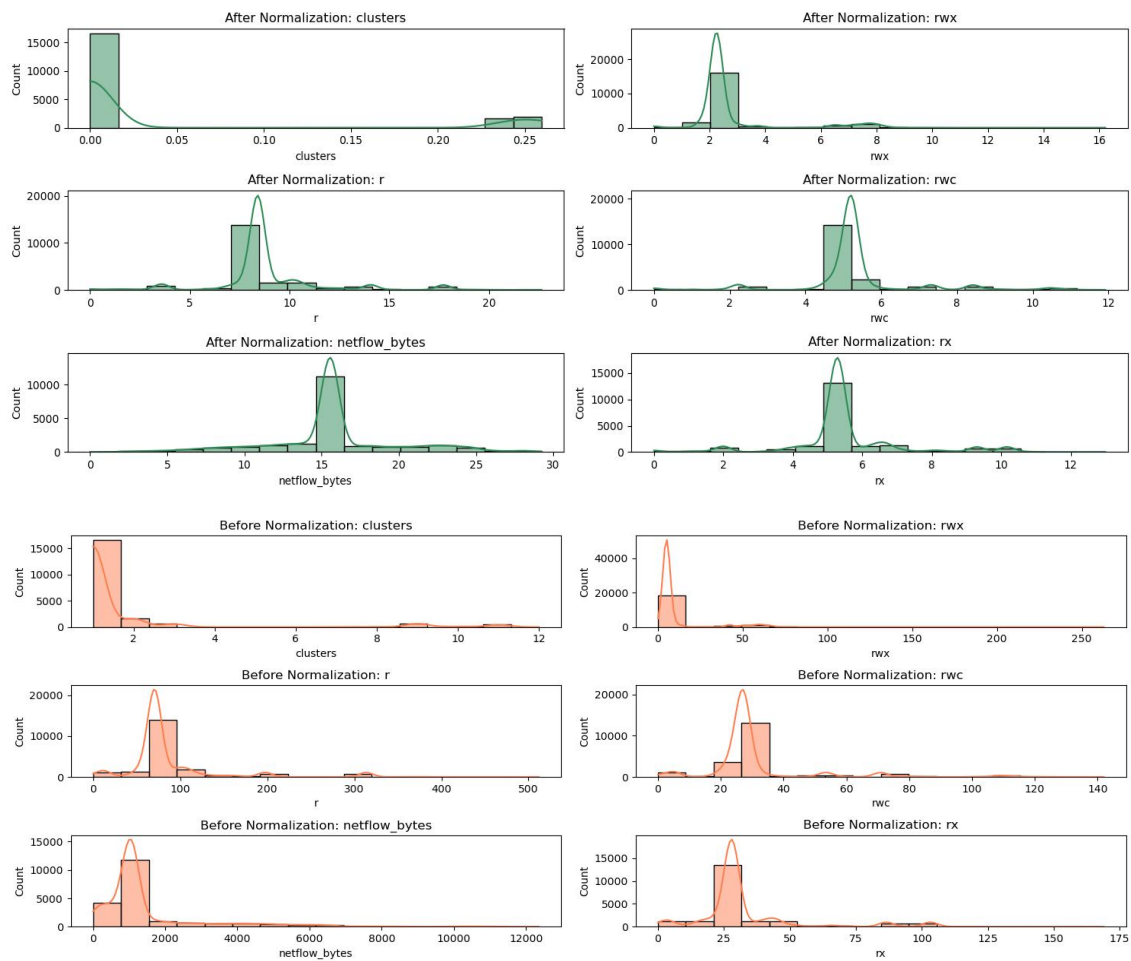

**Figure 10.** Numerical features transformations

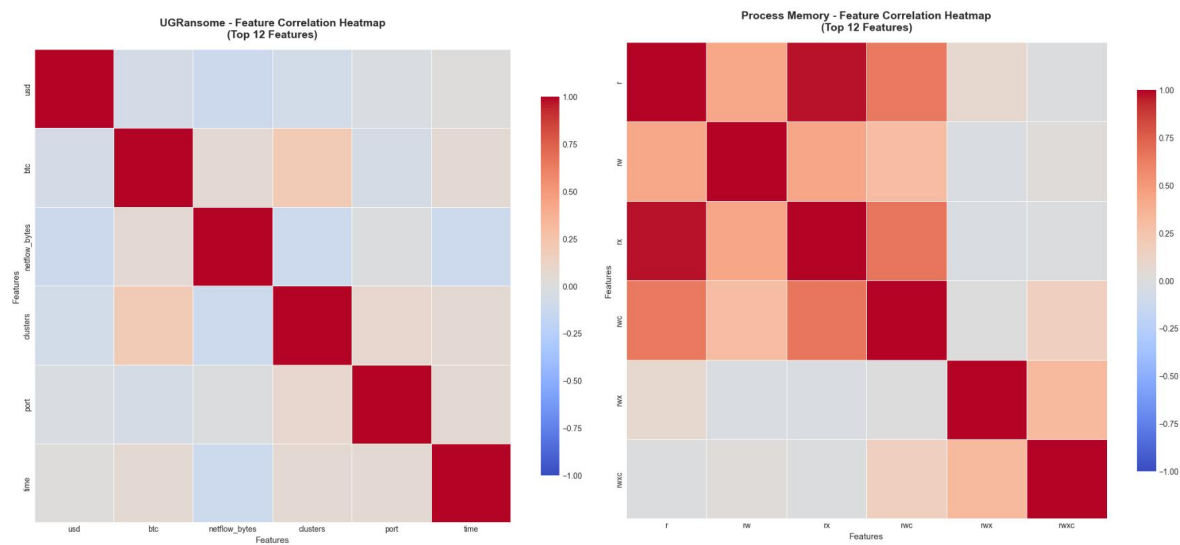

**Figure 11.** Normalized features

In the UGRansome dataset, traffic-based variables such as netflow_bytes and clusters formed strong positive correlation clusters, suggesting shared behavioural trends during ransomware transmission. In the PM dataset, correlations among rw, rwx, and rwxc reflected interdependent memory-access permissions that change under ransomware execution. The heatmap directly supports the research question by identifying feature groups that co-vary in ransomware cases, guiding which features may jointly influence the target label. A composite violin plot (Figure 12) displays all normalized and scaled features for both datasets after transformation and Min-Max scaling to the [0, 1] range. The UGRansome dataset shows 6 features while Process Memory displays 8 features, with the red dashed line marking the 0.5 midpoint. This visualisation addresses the research question by demonstrating how features behave after standardisation. While UGRansome shows broader feature diversity across network patterns, Process Memory shows concentrated clustering around specific memory access permissions. The transformation from highly skewed distributions (btc: 10.82, rw: 15.47) to near-symmetric normalized features prove reliable ransomware detection and improved LLM convergence. In real-world cybersecurity contexts, data normalization and skewness reduction play a crucial role in improving the quality, reliability, and interpretability of ransomware analytics. Cyber datasets such as process memory logs and network traffic captures are often highly skewed. By applying transformations such as logarithmic, square root, or Box-Cox, extreme outliers are compressed, and variable scales are stabilised. This prevents machine learning models like (Bert, RoBERTa, DeBERTa) from being dominated by rare, and high-magnitude observations (de-Marcos et al., 2025). This transformation results in more balanced learning across features, improves model accuracy and convergence rates, and reduces bias towards frequent or extreme values.

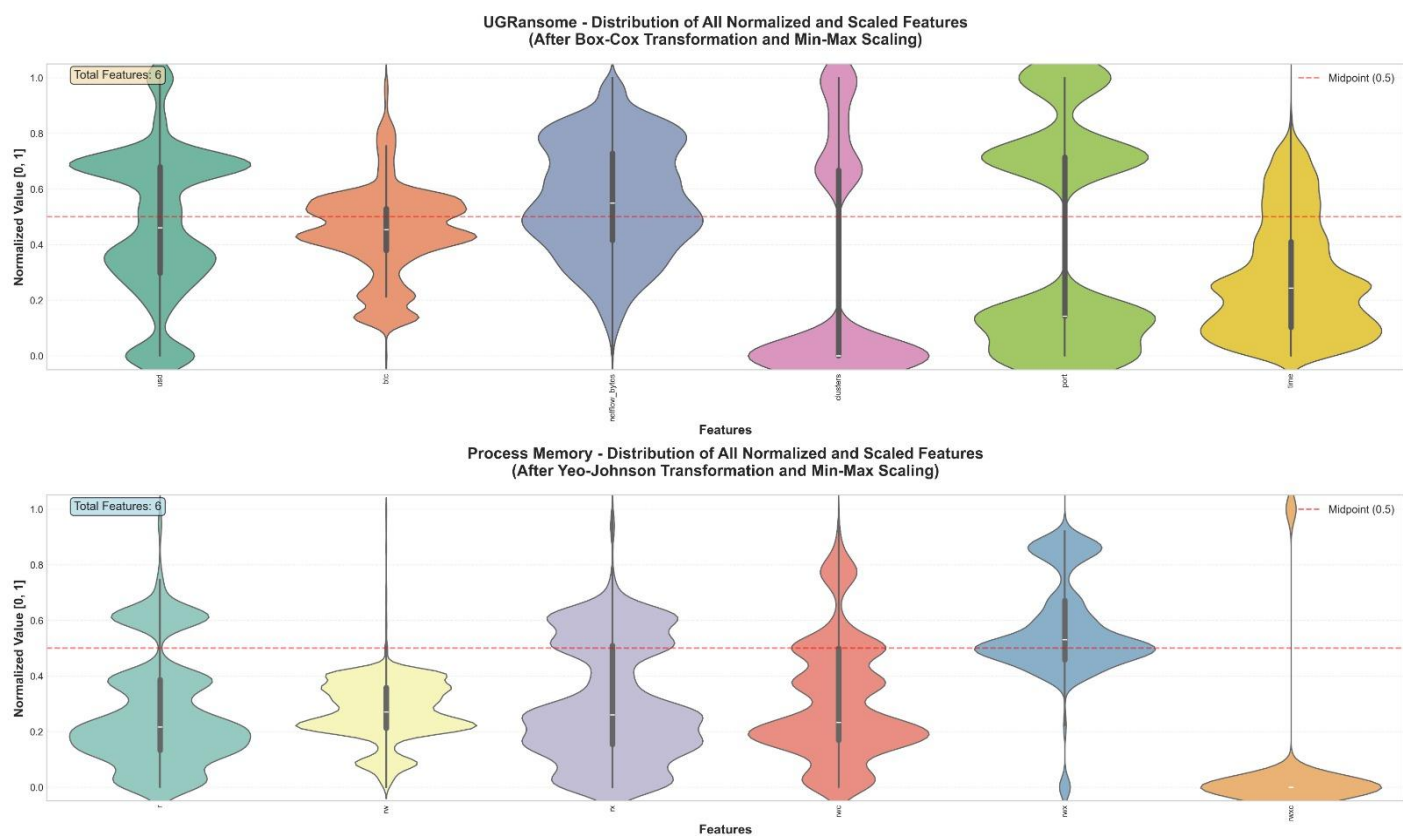

**Figure 12.** Distribution of normalized features

**5.2. Large language models**

Large Language Models (LLMs) are large-scale, pretrained neural language models built primarily on the Transformer architecture that learn contextual token representations through self-attention mechanisms and large-scale pretraining on massive text corpora (Ferraris et al., 2025). They obtain advanced capabilities through scaling of parameters, training data, and computational resources, combined with specific pretraining objectives tied to encoder/decoder design choices (Ferraris et al., 2025; Atkinson, 2024). Three unique LLMs were used: BERT (Bidirectional Encoder Representations from Transformers), RoBERTa (Robustly Optimized BERT Pretraining

Approach), and DeBERTa (Decoding-enhanced BERT with Disentangled Attention). These models are pretrained transformer-based LLMs, which are initially trained on vast amounts of text data to learn general language patterns before being adapted to specific tasks. They rely on a mechanism called self-attention, which allows them to understand the relationships between words in a sentence by considering the full context, both before and after each word. This helps the model grasp meaning more accurately than traditional methods. Once pretrained, these models can be fine-tuned for specific applications, such as classifying text into categories (Gaber et al., 2025). In our case, they are fine-tuned to distinguish between benign text and text associated with ransomware, making them highly effective for cybersecurity-related classification tasks. These models were selected because they offer strong performance and adaptability for text classification (Gaber et al., 2025; Ferraris et al., 2025; Atkinson, 2024). Transformer-based LLMs like BERT, RoBERTa, and DeBERTa consistently outperform traditional machine learning methods by capturing deep contextual relationships in text (Ferraris et al., 2025; Atkinson, 2024). Their bidirectional design allows each word to be understood in relation to the entire sentence, improving accuracy. Since they are pretrained on large datasets, they can be fine-tuned effectively even with smaller samples like the 2,500 used in this study (Wang et al., 2025). Comparing these three models also helps assess how architectural enhancements, such as RoBERTa's extended training or DeBERTa's disentangled attention, impact classification results.

## 6. Model Training

The fine-tuning process used a pretrained BERT-base-uncased model which has around 110 million parameters to classify ransomware related text sequences as Benign or Ransomware. The RoBERTa-base model which has about 125 million parameters was applied in the same way. Deberta-v3-small was chosen due to its smaller parameter size (around 44 million) to speed up the model training time. A random subset of approximately 2,500 samples was selected from the pre-processed dataset, with labels mapped numerically (Benign=0, Ransomware=1). The data was stratified and split into training (80%), validation (10%), and test (10%) sets. Each text sequence was tokenized using the BERT tokenizer, padded or truncated to a maximum length of 128 tokens, and converted into PyTorch tensors (*input_ids and attention_mask*). A custom Ransome Dataset class facilitated batch processing via Data Loader with a batch size of 8. The model was fine-tuned for one epoch using the *AdamW optimizer* with a learning rate of 3e-5 and a linear learning rate scheduler. Training and validation were monitored using accuracy and F1-score, and the final model and tokenizer were saved for future inference. The evaluation of BERT, RoBERTa, and DeBERTa showed unique performance characteristics and decision making approaches for ransomware detection. Each model demonstrated strengths and limitations when processing.

### 6.1. Model interpretation

BERT's metrics are very close to RoBERTa's but slightly lower in balanced performance (F1). However, it has a higher ROC-AUC score than RoBERTa. This high ROC score indicates good inherent power to distinguish between benign and ransom classes, with a precision of 0.821 which is excellent for not generating excessive numbers of false positives. Its weakness lies in all the other metrics such as accuracy, recall and F-1 score, where RoBERTa slightly outperforms it. RoBERTa has the highest Accuracy, Recall, and F1-Score of the three models. Its Precision is also virtually tied for the highest. RoBERTa is the most consistent and well-balanced model, as it performs with the best accuracy which allows it to find true threats and avoiding false alarms from its F-1 score. For a cybersecurity organization wanting a single, reliable model that performs well across the board without a major weakness, RoBERTa is the recommended choice. Its high F1-score indicates it is the most robust for general deployment. DeBERTa is the model with the highest precision and ROC-AUC score, but with the lowest accuracy, recall and F-1 score. This model is the most confidence inspiring choice, as when it flags ransomware it has the highest confidence of 83.7%. Its main drawback is due to its low recall score of 0.772 which means it will miss more actual ransomware attacks that the other models, about 23% compared to 21% for the other models. From these insights we can see that RoBERTa and BERT are tuned for a balance or slightly higher recall, while DeBERTa offers the highest precision. The AUC score of DeBERTa suggests that it has the greatest inherent capability. Its lower Recall might be improved without sacrificing too much Precision by lowering its classification threshold, potentially allowing it to match or exceed the others. For a default system, RoBERTa is the best choice, as it provides the highest overall performance and reliability. For maximising trust in alerts, DeBERTa is the best due to its high precision score and makes it operationally efficient for analyst. BERT remains a very strong and valid option, but is still outperformed in practical metrics.

# 7. Results

*7.1. What are key features from UGRansome and PM datasets crucial for ransomware classification?*

The UGRansome and Process Memory (PM) datasets each provide complementary insights into ransomware classification. The UGRansome dataset focuses on network-level behaviours, capturing the external communications of ransomware (Fan et al., 2025). This dataset comprises a mixture of benign and malicious network behaviours including a number of categories like UDP scans, port scans, SSH connections and malicious behaviours like botnets, spam and denial of service (DoS) attacks (Mohamed et al., 2025; Su, 2025). The important features of the UGRansome dataset include bytes transferred, malicious IP addresses, connection duration and protocol flags giving flow-oriented indications of communication concerns. These features describe how the ransomware communicates over the network, how the data is transferred to external servers and whether unusual traffic load or repeated communication attempts with known malware is produced. Such behaviours must be investigated in order to classify ransomware at the early infiltration and propagation stages. On the contrary, the Process Memory (PM) dataset gives host-level behavioural features which describe the internal behaviours of ransomware once the software has infiltrated a victim's platforms (Singh et al., 2023; Mokoma and Singh, 2025). This dataset captures memory usage, exploitation of access privileges, the frantic traces of execution of the ransomware, an all of which are manifestations of how the ransomware is manipulating the local environment in order to encrypt data or adjust system configurations (Mahboubi, et al., 2025). This dataset captures permission types like read (r), write (w) and execute (x) combinations (e.g., rwx or rwxc). This gives information as to how ransomware processes gain and promote privileges. These messages are valuable as they emanate from the sandboxed environments which execute these processes and are useful for identifying key behaviours like encryption attempts and abnormal access of memory as well as the escalation of privileges which are occurring simultaneously in actual attacks (Mokoma and Singh, 2025; Singh et al., 2023). Therefore, the UGRansome and Process Memory (PM) datasets are comprehensive sources basis of ransomware detection. UGRansome is able to find the external indicators of compromise from the traffic flows, while PM finds the internal fingerprints of operations on memory usage and executions of processes. By combining the network view and the system view this classification framework is capable of detecting some aspect of the ransomware over multiple stages of the lifecycle, from initial infection and command and control communication to the point of active encryption and execution.

*7.2. How do XAI techniques like LIME and SHAP improve the interpretability of ransomware classification models?*

Explainable AI (XAI techniques like LIME and SHAP fundamentally enhance the interpretability of ransomware classification models by transforming them from opaque black boxes into transparent decision-support tools. LIME runs locally and is used to explain predictions by identifying which features are identify as suspicious network port or bitcoin transactions (Vikas Hassija, 2023). SHAP provides a global, model-agnostic perspective rooted in game theory, calculating the fair contribution of each feature to the model's predictions across all samples. The analysis of BERT, RoBERTa, and DeBERTa shows the power of XAI techniques in models interpretability. Being able to interpret the training results into an understandable output is the key for actionable insights to improve and make decisions based on results. For instance, LIME revealed that DeBERTa correctly flags ransomware by focusing on the absence of normal file operations (rwxc_bin_0: -0.1769) combined with high read-write activity, a clear signature of file encryption. SHAP quantified DeBERTa's strategic reliance on Bitcoin transactions (btc_bin_3: 0.0203 impact) as its primary global signal, which aligns perfectly with the domain knowledge that ransomware involves financial demands. Between LIME's local clarity and SHAP's globally consistent explanations, the combined use of both methods provides substantial analytical advantages and increases confidence in the model's reasoning. This dual approach enhances trust in the predictions made for ransomware threats and supports the development of more accurate and interpretable models for detecting similar emerging attacks in the future.

*7.3. How does the proposed framework (LLM-XAI) perform compared to traditional methods?*

The proposed LLM-XAI framework has been shown to outperform traditional machine learning models including K-Nearest Neighbour (K-NN), Reinforcement Learning (RL) and Recurrent Neural Networks (RNN), through pure performance and interpretability. Traditional models like K-NN rely mainly on distance between similar features and cannot capture contextual relationships well in a complex dataset such as one found in ransomware.

*Model evaluation and results comparison*

### DEBERTA – Confusion Matrix

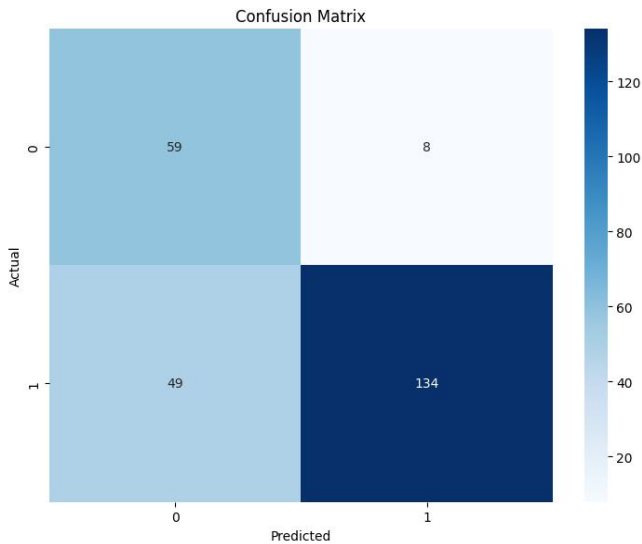

### RoBERTa – Confusion Matrix

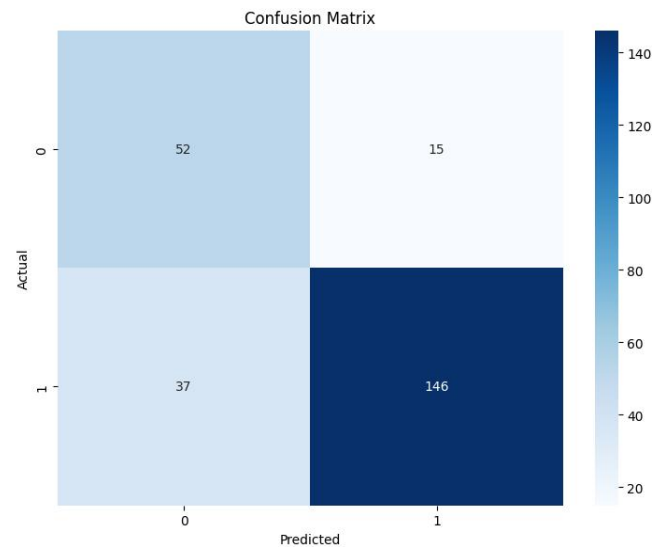

### BERT – Confusion Matrix

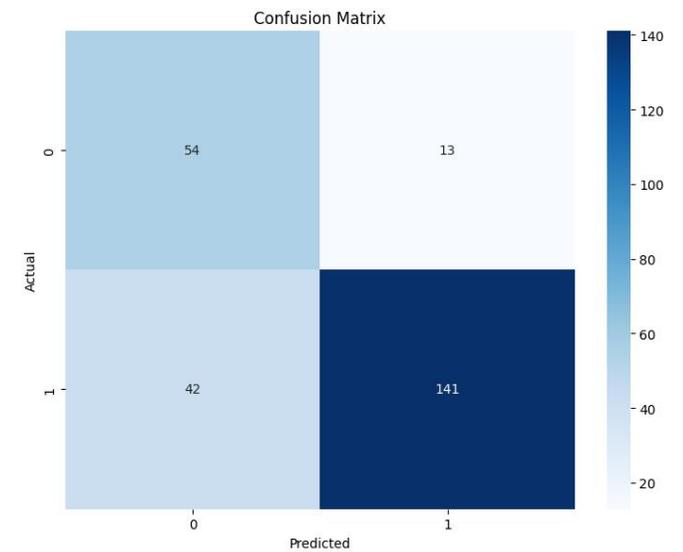

From 250 samples, the DeBERTa model correctly identified 59 harmless files and 134 ransomware files, giving an overall accuracy of 77.2%. However, the model did make errors by incorrectly flagging 8 safe files as ransomware (false positives) and missed 49 actual ransomware attacks (false negatives). The model achieved 73.3% precision, meaning when it predicts ransomware, it's usually right. The recall was 94.3%, indicating it successfully caught about 94% of all ransomware instances. The false positive rate is the most concerning factor, as it means there are more false alarms. But a low False Negative score indicate that there will be little ransomware attacks that won't be recognised.

The models confusion matrix shows a mixed result. The model correctly classified 146 samples as ransomware (True Positives), while only having 37 False positives, which indicate a very good recall of 90% to ransomware samples. A total of 52 True negatives with only 15 False Negatives show that the model does well to predict Benign records with a precision of 79.7%. The model's main downside is the amount of False Positives (37), and indicate that the model is overly cautious toward ransom samples. It has learned to detect malicious patterns well but sometimes misinterprets unusual but benign activity as malicious. The RoBERTa confusion matrix compares closely to the BERT matrix, but with more TP and less FP which is a good indication of better model performance.

The BERT model showed a very useful profile for security application. From the 250 samples tested, the model could identify 141 of Ransomware as True positives, which indicate a very high (around 90%) recall on ransomware detection, while only getting 13 False Negatives. The main area for improvement is the high number of False Positives (42), which results in a moderate Precision of 82.3%. This suggests that the model is being somewhat "overly cautious." It has learned to detect malicious patterns well but sometimes misinterprets unusual but benign activity as malicious.

| DEBERTA – ROC Curve | RoBERTa – ROC Curve | BERT – ROC Curve |
|---|---|---|
| 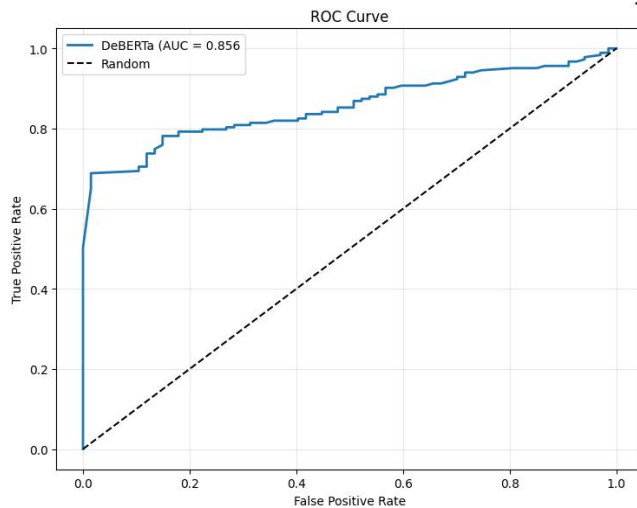 | 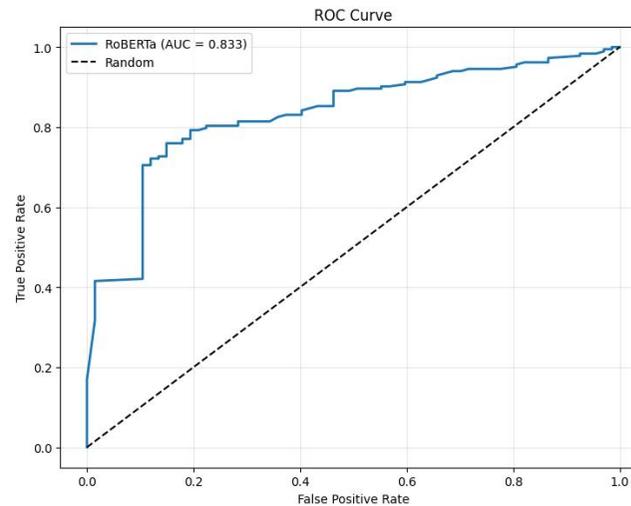 | 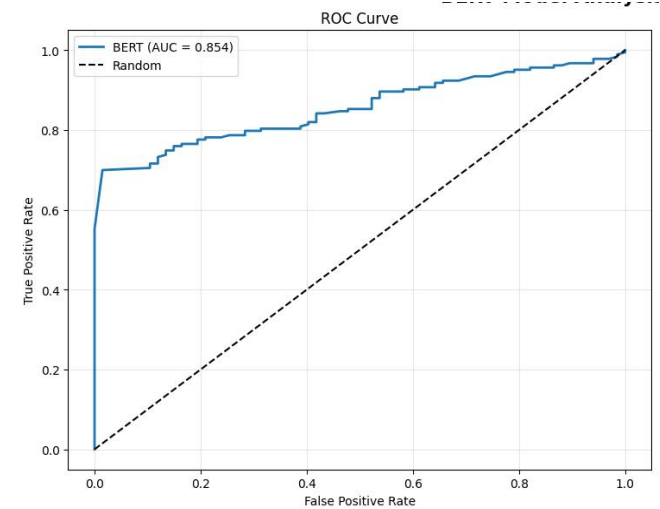 |
| The DeBERTa model achieves an AUC (Area Under the Curve) of 0.856, indicating strong discriminative ability. An AUC of 0.856 means the model correctly ranks a random positive instance higher than a random negative instance 85.6% of the time. The curve rises steeply initially and stays well above the diagonal dashed line (which represents random guessing with AUC = 0.5). This demonstrates that the model significantly outperforms random classification models. DeBERTa's enhanced performance can be attributed to its novel architecture, which features disentangled attention and an enhanced mask decoder. This model has the highest AUC score of all 3 models, and is because of the very high recall of 94% with slightly better or similar scores in precision, recall and F-1 score. | The RoBERTa ROC curve demonstrates excellent classification performance with an AUC of 0.833, indicating strong discriminative ability between ransomware and benign samples. The curve's shape shows that RoBERTa achieves high true positive rates while having consistently low false positive rates across various classification thresholds. An AUC of 0.833 suggests the model has about 83% probability of correctly ranking a random ransomware sample higher than a randomly chosen benign sample. This performance level indicates RoBERTa is highly effective for ransomware detection. | The BERT ROC curve demonstrates strong classification performance with an AUC of 0.854, indicating good discriminative ability between ransomware and benign samples. The curve stays well above the random classifier diagonal line throughout most thresholds, indicating BERT can effectively distinguish between the two classes. An AUC of 0.854 means there's approximately an 85% probability that BERT will correctly rank a randomly selected ransomware sample higher than a randomly selected benign sample. While slightly higher than RoBERTa's AUC of 0.833, this performance level represents solid classification capability for ransomware detection. |

| DeBERTA – Attention Weights | RoBERTa – Attention Weights | BERT – Attention Weights |
|---|---|---|
| 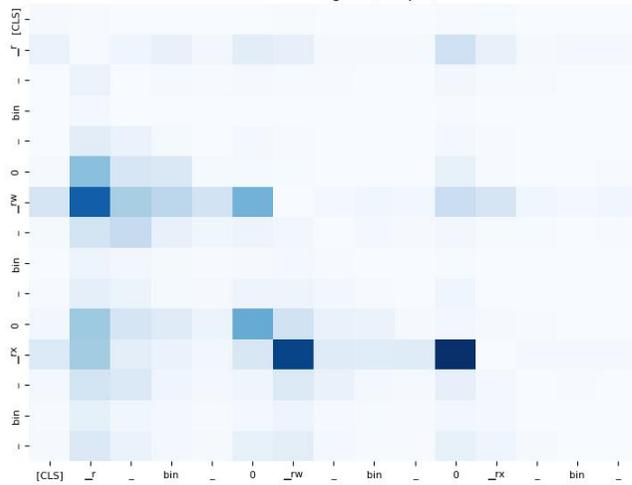 | 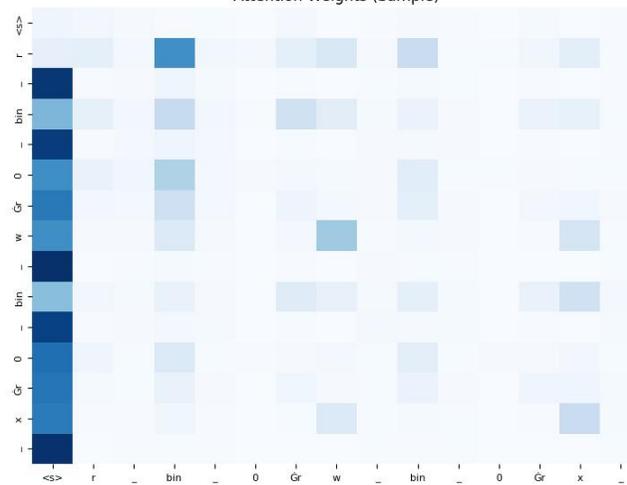 | 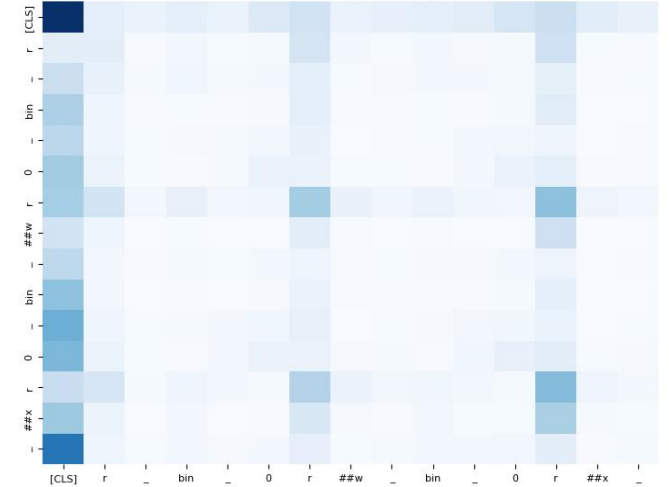 |
| The attention weights visualization for the DeBERTa model reveals a sophisticated and highly contextual pattern of focus. Unlike simpler models, DeBERTa demonstrates an advanced, non-uniform strategy, dynamically allocating attention across different bin tokens and specific operations (rx, rw). The model shows varying levels of attention to different bin tokens. Crucially, it places significant emphasis on tokens that are contextualized by operation codes like rx (read-execute) and rw (read-write). This confirms the success of the feature transformation and tokenization process. The model is effectively interpreting this engineered "language" of system behavior. DeBERTa shows a more wildly spread attention across all of the bin tokens, which is unlike the RoBERTa and BERT models, which focus more on the initial bins. | The plot reveals a highly focused and context-aware pattern, indicating that the model successfully identifies and prioritises specific, critical features from the tokenized numerical data. The deep blue colouring indicates that RoBERTa places nearly all its focus on these initial tokens, particularly the first token which likely links to the <s> classification token. The presence of multiple attention heads in RoBERTa's architecture allows it to capture different types of relationships simultaneously. While this view may aggregate them, it confirms that the model is building a contextual understanding by relating the currently viewed token to these specific, high-importance bins elsewhere in the sequence. While this concentrated attention pattern suggests that RoBERTa has a strong performance, it may also suggest the model is less able to feature variations and could be more vulnerable to attacks that affect the early tokens. | The visualization confirms that the model is dynamically and contextually focusing on specific, discretized features within the input sequence to make its classification, enhancing the transparency and interpretability of the ransomware detection system. This concentration demonstrates that the model is not treating all features equally. It successfully identifies and pays more attention to a select group of discretized features that are most indicative of ransomware behavior for this specific sample. The highly attended tokens (visible as bright lines in the plot) represent the feature bins that the model found most salient for its decision. This ability to focus on specific tokens over others is a core strength of the transformer architecture. |

| DEBERTA – Performance Metrics | RoBERTa – Performance Metrics | BERT – Performance Metrics |
|---|---|---|
| 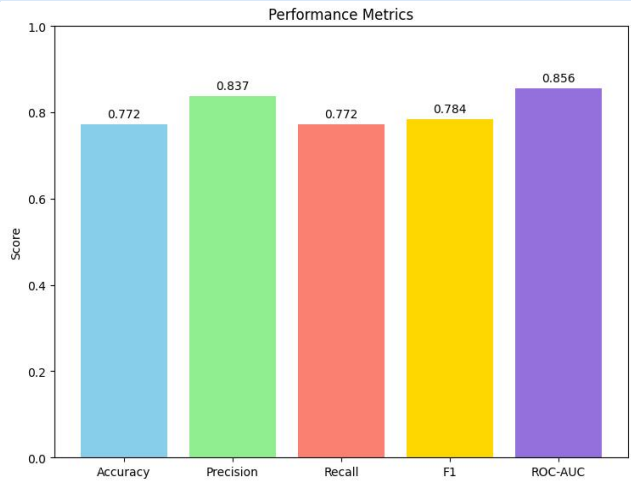 | 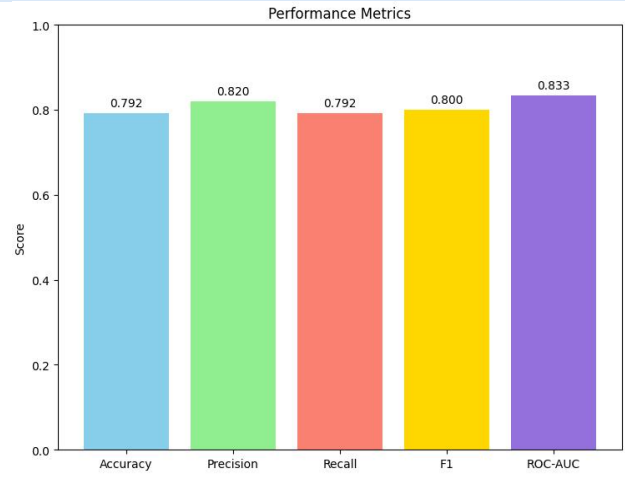 | 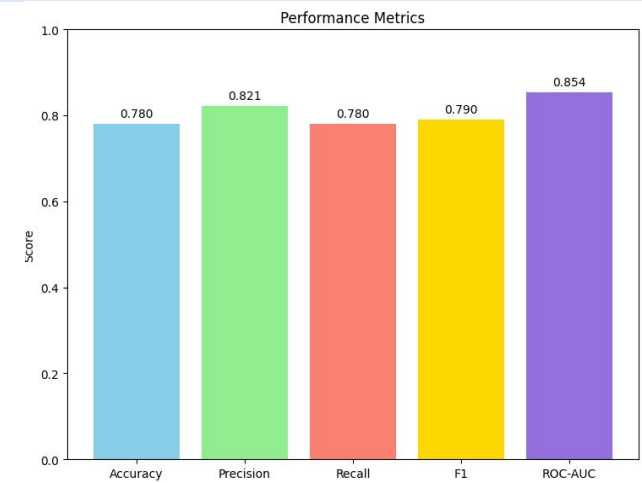 |
| The DeBERTa performance metrics chart shows strong overall results for the model across all evaluation measures. The model achieved 77.2% accuracy, indicating it correctly classified approximately 4 out of 5 ransomware samples. The precision of 83.7% demonstrates that when the model predicts ransomware, it's correct 83% of the time, which helps reduce false alerts. The recall of 77.2% shows the model successfully detects 77% of actual ransomware events, though the 23% miss rate could be concerning in critical security scenarios. The F1-score of 78.4% represents a good balance between precision and recall. Most notably, the ROC-AUC score of 85.6% indicates excellent discriminative ability, showing the model can effectively distinguish between benign and malicious samples. These metrics collectively demonstrate that DeBERTa provides reliable ransomware detection performance suitable for real-world cybersecurity applications, though the recall could be improved to reduce the risk of undetected attacks. | The RoBERTa performance metrics revealed a good performance. The model achieved 79.2% accuracy, which is good and slightly better than BERT's 78%, but better than DeBERTa's 77.2%. The precision is also significantly at 82%, indicating that when RoBERTa predicts ransomware, it's correct about 4 out of every 5 times. The recall of 79.2% shows the model detects approximately 79% of actual ransomware instances, missing about 21% of attacks. The F1-score of 80% reflects the good balance between precision and recall, heavily impacted by the high precision. The ROC-AUC score of 83.3% is quite strong, suggesting the model has good discriminative ability across different thresholds despite the classification issues. Overall the RoBERTa performs slightly better than the BERT and DeBERTa in accuracy, except for the slightly lower ROC-AUC score with a 2.1% difference in the models. | The BERT performance metrics show near identical results to the RoBERTa performance. With 78% accuracy and the same strong precision of 82.1%. The model achieves 78% recall, meaning it detects about 78% of actual ransomware instances while missing 22% of attacks. The high precision indicates that when BERT flags something as ransomware, it's correct about 4 out of every 5 times, resulting in few false alarms. The F1-score of 79% reflects this good precision-recall balance. However, the ROC-AUC of 85.4% demonstrates good discriminative ability, suggesting the model can distinguish between classes well but may need minor threshold adjustment. BERT as the "smallest" model performed very well overall, with performance metrics very close to both the DeBERTa and RoBERTa, showing that this model is well capable in ransomware classification. |

Selected LLMs performed well on simple, low dimensional datasets but when the datasets become more non-linear or when we see an identical behaviour arising out of different features, as would happen with network traces or memory traces, they fail to generalize. Reinforcement leaning can factually improve detection policies, but is resource hungry, often requiring environment simulation and an extensive loop of reward feedback which will not work in practice in real time detection of a ransomware type incident. Similarly, while RNNs are capable of modelling sequential dependencies, they rely on the heavy handcrafting of feature engineering and are not as effective in modelling the deep semantic and structural patterns of mind exhibited by ransomware communications and execution behaviours (Alzakari, et al., 2025). In contrast, the proposed LLM-XAI framework based on BERT, RoBERTa and DeBERTa leverage sequential and contextual relationships, using self-attention methods. This allows for learning the relationships that exist among the features on a network and host level, relationships which are largely ignored by traditional algorithms. Furthermore, incorporating Explainable AI techniques such as LIME and SHAP strengthens the advantages offered by these advanced models. By revealing the contribution of each feature to the final classification, these methods add essential transparency and make it possible to understand why the model selects one decision path over alternative roots in each case. Empirically LLMs have been shown to have greater accuracy, F1 scores, and ROC-AUC, representing a more stable and reliable performance across datasets like the UGRansome and PM. The attention maps and the feature importance figures can lay bare the workings of the model which traditional algorithms do not exhibit with the same clarity, and this allows for a greater level of trust in the system by the analyst and compliance with explainable forms in the provision of cybersecurity functionalities. Overall, the LLM-XAI framework, delivers superior detection accuracy, deeper contextual understanding and higher interpretability compared to traditional approaches. It improves the precision of ransomware classification and provides meaningful, explainable insights into why certain behaviours are actually malicious, leading to a much more transparent, and trustworthy detection system.

*7.4. What are the limitations of applying XAI techniques (LIME Vs. SHAPE) to ransomware classification using LLMs (BERT, RoBERTa, and DeBERTa)?*

Models like BERT do not process raw tokens but rather their dense embeddings. Perturbing a token creates a new, arbitrary embedding that may not lie in a semantically meaningful region of the model's latent space. This can lead to explanations based on nonsensical or out-of-distribution data points, reducing their reliability. Both techniques require running the model hundreds or thousands of times for a single explanation. Given that these are large, computationally intensive models, generating explanations in real-time is often impractical and this limits their use to offline analysis and spot-checking. Our DeBERTa model had the highest precision and is deemed as the most confidence inspiring choice. However, its low recall means that the model misses many of the true positives. A key limitation of XAI is that it only explains the predictions that are made. For DeBERTa the explanation will be very confident and correct for the alerts it raises, but the XAI cannot show or explain why it missed 23% of the ransomware that RoBERTa caught.

## 8. Conclusion

We propose an approach that transformed numerical features into discrete tokens, fine-tuned three transformer architectures, and employed LIME and SHAP to explain their prediction pathways. The evaluation results show that all three LLMs performed robustly, each demonstrating distinct strengths. RoBERTa emerged as the most balanced model, achieving the highest overall accuracy and F1-score, making it the preferred choice for a reliable, general-purpose detection system. DeBERTa delivered the highest precision and ROC-AUC score, making it particularly well suited for operational environments where minimizing false alarms is critical. BERT offered a strong and stable baseline, illustrating that even foundational transformer models remain highly competitive in structured cybersecurity tasks. The XAI analysis using LIME and SHAP confirmed that the models relied on meaningful and domain-plausible indicators, such as network port behaviors (e.g., port_bin_1), Bitcoin-related activity (e.g., btc_bin_3), and the absence of expected file-operation patterns (e.g., rwxc_bin_0). These insights demonstrate that the models are not functioning as opaque black boxes; rather, they are learning interpretable and cybersecurity-relevant representations that align with established ransomware behaviors. In summary, transformer-based LLMs such as BERT, RoBERTa, and DeBERTa can be effectively adapted for ransomware classification when numerical telemetry is tokenized into an LLM-compatible format. The integration of XAI techniques provides essential transparency, enabling analysts to understand and trust the reasoning behind model predictions. This combination of high predictive performance, interpretability, and operational reliability is vital for deploying AI systems in high-stakes cybersecurity environments.

**Appendix A**

In the table below you will find the basic statistics for all of the numerical columns in the datasheets.

| Numerical column | count | mean | std | Min | 1% | 25% | 50% | 75% | max | mode |
|---|---|---|---|---|---|---|---|---|---|---|
| r | 20186.0 | 85.836719 | 57.310929 | 0.0 | 3.0 | 71.0 | 71.0 | 71.0 | 512.0 | 71.0 |
| rw | 20186.0 | 97.090756 | 135.135509 | 1.0 | 4.0 | 73.0 | 73.0 | 73.0 | 7217.0 | 73.0 |
| rx | 20186.0 | 32.888338 | 20.474955 | 0.0 | 1.0 | 28.0 | 28.0 | 28.0 | 169.0 | 28.0 |
| rwc | 20186.0 | 30.025612 | 16.837615 | 0.0 | 0.0 | 27.0 | 27.0 | 27.0 | 142.0 | 27.0 |
| rwx | 20186.0 | 10.207371 | 15.460123 | 0.0 | 0.0 | 5.0 | 5.0 | 5.0 | 263.0 | 5.0 |
| rwxc | 20186.0 | 0.086694 | 1.223344 | 0.0 | 0.0 | 0.0 | 0.0 | 0.0 | 37.0 | 0.0 |
| usd | 20186.0 | 8750.124839 | 19567.170009 | 1.0 | 1.0 | 3044.5 | 3044.5 | 3044.5 | 126379.0 | 3044.5 |
| btc | 20186.0 | 22.187457 | 72.989503 | 1.0 | 2.0 | 13.0 | 13.0 | 13.0 | 1864.0 | 13.0 |
| netflow_bytes | 20186.0 | 1531.850045 | 1684.926056 | 1.0 | 23.0 | 1038.5 | 1038.5 | 1038.5 | 12360.0 | 1038.5 |
| clusters | 20186.0 | 1.696968 | 2.150578 | 1.0 | 1.0 | 1.0 | 1.0 | 1.0 | 12.0 | 1.0 |

Below is an example of data after preprocessing, please note that the "label" has not yet been removed.

| Row index | Text (tokenised) | label | source |
|---|---|---|---|
| 11919 | r_bin_2 rw_bin_2 rx_bin_2 rwc_bin_2 rwx_bin_1 … | Ransomware | network_traffic |
| 17614 | r_bin_2 rw_bin_2 rx_bin_2 rwc_bin_2 rwx_bin_1 … | Benign | network_traffic |
| 11526 | r_bin_2 rw_bin_2 rx_bin_2 rwc_bin_2 rwx_bin_1 … | Benign | network_traffic |
| 1815 | r_bin_0 rw_bin_0 rx_bin_0 rwc_bin_0 rwx_bin_0 … | Ransomware | process_memory |
| 18925 | r_bin_2 rw_bin_2 rx_bin_2 rwc_bin_2 rwx_bin_1 … | Benign | network_traffic |

**Datasets and code**

https://www.kaggle.com/datasets/nkongolo/ugransome-dataset/data

https://www.kaggle.com/code/thashannaick/ransomware-detection-using-llm-and-xai-techniques

```python
#=========================================================================================
=====#
# Lime and Shap for DEBERTA   - PM Ransomware
#=========================================================================================
=====#

pipe = pipeline(
    "sentiment-analysis",
    model=modelDEBERTAPM,          # your trained DeBERTa/BERT/RoBERTa model
    tokenizer=tokenizer,
    return_all_scores=True
)

# --- Create a simple prediction function for LIME using pipeline ---
def predict_proba(texts):
    all_probs = []
    for t in texts:
        pred = pipe(t)  # returns [[{label, score}, {label, score}]]
        pred = pred[0]  # unwrap the inner list
        # sort by label to ensure consistent order
        probs = [p['score'] for p in sorted(pred, key=lambda x: x['label'])]
        all_probs.append(probs)
    return np.array(all_probs)
```